\documentclass{aa}  

\usepackage{graphicx}
\usepackage{txfonts}

\usepackage{graphicx}
\usepackage{subfig}
\usepackage{nth}
\usepackage{longtable}
\usepackage{color}
\usepackage{hyperref}
\usepackage{txfonts}
\usepackage{natbib}
\bibpunct{(}{)}{,}{a}{}{,}
\newcommand{\kms}{\mbox{${\rm km\,s}^{-1}$}}

\newcommand{\teff}{\mbox{$T_{\rm eff}$}}
\newcommand{\logg}{\mbox{$\log g$}}

\newcommand\T{\rule{0pt}{2.2ex}}

\begin{document}

   \title{Signs of strong Na and K absorption in the transmission spectrum of WASP-103b}


   \author{M. Lendl
          \inst{1,2}
           \and
           P.~E. Cubillos\inst{1}
           \and
           J. Hagelberg\inst{3,4}
           \and
           A. M\"uller\inst{2}
           \and
           I. Juvan\inst{1,5}
           \and 
           L. Fossati\inst{1}
          }

   \institute{Space Research Institute, Austrian Academy of Sciences, Schmiedlstr. 6, 8042 Graz, Austria\\
              \email{monika.lendl@oeaw.ac.at}
         \and
         Max Planck Institute for Astronomy, K\"onigstuhl 17, 69117 Heidelberg, Germany 
         \and 
         Institut de Planétologie et d’Astrophysique de Grenoble, Universit\'e Grenoble Alpes, CS 40700, 38058 Grenoble Cédex 9, France
         \and
         University of Hawaii, 2680 Woodlawn Dr., Honolulu HI 96822, USA
         \and
         Institut f\"ur Geophysik, Astrophysik und Meteorologie, Karl-Franzens-Universit\"at, Universit\"atsplatz 5, 8010 Graz, Austria\\
         }

   \date{}

 
  \abstract
  {Transmission spectroscopy has become a prominent tool for characterizing the atmospheric properties
  on close-in transiting planets. 
  Recent observations have revealed a remarkable diversity in exoplanet spectra, which show absorption signatures of Na, K and $\mathrm{H_2O}$, 
  in some cases partially or fully attenuated by atmospheric aerosols. Aerosols (clouds and hazes) themselves have been detected in the
  transmission spectra of several planets thanks to wavelength-dependent slopes caused by the particles' scattering properties.}
   {We present an optical 550 -- 960~nm transmission spectrum of the extremely irradiated hot Jupiter WASP-103b, one of the hottest (2500~K) 
  and most massive (1.5~$M_J$) planets yet to be studied with this technique. WASP-103b orbits its star at a separation of less than 1.2 times the Roche limit
  and is predicted to be strongly tidally distorted.}
   {We have used Gemini/GMOS to obtain multi-object spectroscopy 
   throughout three transits of WASP-103b. We used relative spectrophotometry and bin sizes between 20 and 2~nm to infer the planet's transmission spectrum.}
   {We find that WASP-103b 
  shows increased absorption in the cores of the alkali (Na, K) line features. We do not confirm the presence of any strong 
  scattering slope as previously suggested, pointing towards a clear atmosphere for the highly irradiated, massive exoplanet WASP-103b.
  We constrain the upper boundary of any potential cloud deck to reside at pressure levels above 0.01~bar.
  This finding is in line with previous studies on cloud occurrence on exoplanets which find that clouds dominate the transmission spectra
  of cool, low surface gravity planets while hot, high surface gravity planets are either cloud-free, or possess clouds located
  below the altitudes probed by transmission spectra.}
   {}


   \maketitle
%

\section{Introduction}

Due to their high temperatures and extended atmospheres, hot Jupiters (i.e, gas giants at orbital periods of a few days) are ideal laboratories 
for studying the properties of exoplanet atmospheres. Transmission spectroscopy probes the 
planetary atmosphere's composition and structure by measuring the spectrally resolved absorption during transit. 
A growing sample of hot-Jupiter transmission spectra from ground and space is revealing a remarkable diversity in spectral signatures. Some planets
show prominent absorption signatures of alkali (Na, K), and water \citep[e.g.,][]{Charbonneau02,Deming13,Nikolov16,Sedaghati16}.
Many objects however show attenuated features \citep[e.g.,][]{Chen17,Sing15,Sing16}, and in some the elemental or molecular absorptions are entirely 
absent at low spectral resolution \citep{Pont08,Lendl16}.  
Several planets show pronounced slopes of increasing absorption towards shorter wavelengths, as can be explained by scattering on
aerosols with small particle sizes (clouds and hazes) lifted to high altitudes in the planetary atmospheres.
The level of observed attenuation of the absorption features depends on the particles' abundance, size and atmospheric altitude.
The observed slopes stem from scattering on small aerosoles and the slope extension and amplitude depends on the atmospheric temperature 
and the size distribution of the scattering particles \citep{Lecavelier08a}.

The relation between planetary characteristics and the occurrence and properties of aerosols is yet
to be definitely understood. \citet{Stevenson16} and \citet{Heng16} propose that planets with hot atmospheres 
posses less pronounced cloud features, \citet{Stevenson16} also suggests that clouds are more common for low surface gravity 
planets. 

In this work, we present the transmission spectrum of one of the most irradiated transiting hot Jupiters known to date, WASP-103b.
WASP-103b stands out due to its short orbital period of 0.9~d, which is far below
the peak of the hot Jupiter period distribution of about three days \citep[e.g.,][]{Udry03}. Due to tidal dissipation of orbital energy, 
most hot Jupiters are suffering from orbital decay and are spiraling into their host stars \citep[e.g.,][]{Matsumura10,Patra17}. 
WASP-103b is a hot Jupiter in its final stages of orbital decay. Currently, the planetary orbital separation from the host star 
amounts to only 1.16 times its Roche limit, leading to large tidal distortion of the planet \citep{Budaj11,Leconte11}. 
This fact, paired with the planet's extremely high stellar irradiation makes this object 
prone to enhanced mass loss \citep{Lammer03,Vidal03}.

We describe our GMOS observations in Sect. \ref{sec:obs}, and detail our analysis approach, and the correction 
of our measurements for contamination from a nearby star in Sect. \ref{sec:mod}. In Sect. \ref{sec:res}, we present and discuss our results for WASP-103b,
before concluding in Sect. \ref{sec:conc}.

\section{Observations and data reduction}
\label{sec:obs}

\subsection{GMOS-North observations}

We observed WASP-103 throughout three full transits with GMOS at Gemini North during the nights of 27 June 2015, 10 July 2015, 
and 03 May 2016 (all UT, program numbers GN-2015A-Q-63 and GN-2016A-Q-26, PI: Hagelberg). All observations were carried out in multi-object spectroscopy mode 
using the R400 grism and order separation filter OG515. We used custom-cut masks to place slits on the target and two reference
stars. The slit width was set to 10 arcsec for all stars to avoid slit losses due to guiding inaccuracies or seeing variations, and the slit length
was set to 40 arcsec for the target and the brighter reference to allow for a good sky determination. For the fainter reference, the slit
needed to be kept shorter (25 arcsec) to avoid contamination from a nearby star.

\begin{figure}
 \includegraphics[width=\linewidth]{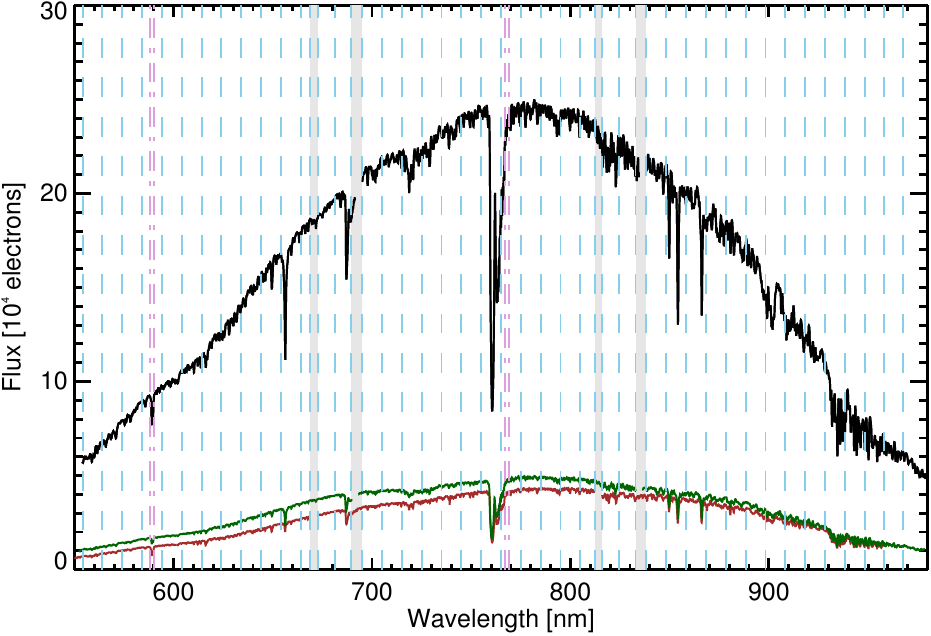}
 \caption{\label{fig:spec}Average spectra of target (top, black) and two reference stars. The 10~nm bins used for spectrophotometric extraction 
 are indicated with blue dashed lines, while the purple dash-dotted lines indicate 5~nm bins centered on the Na and K features. Gaps between
 the GMOS detectors are shaded in gray.}
\end{figure}

Spectral coverage of the target and the brighter reference star is between 550 and 980~nm (top and middle spectra in Fig. \ref{fig:spec}), 
while spectra of the fainter reference star (bottom spectrum in in Fig. \ref{fig:spec}) are slightly offset and cover 530 to 960~nm.
Two amplifiers were used to read each of the three CCDs in 1x1 binning and ``Slow read" mode.
We used custom regions of interest (ROI) to window the detectors and read only the detector areas containing data. 
Sky conditions were clear throughout all three observations, however, the observations on 10 July 2015 were interrupted for 
approx. 22~minutes due to high ambient humidity. We used exposure times of 240~s on 27 June 2015 and 10 July 2015, 
and exposure times of 180~s on 03 May 2016. 

\subsection{Spectral extraction and wavelength calibration}
We carried out the standard calibrations using the GMOS IRAF routines, and also used these routines to calculate an initial
wavelength calibration based on (CuAr) arc frames. The spectra have a pixel scale of 0.68 \AA \, per pixel in the dispersion direction, and a spectral 
resolution of $R\sim1200$.
Cosmic rays were removed using the LA-cosmic routines \citep{vanDokkum01}. 
The resulting 2D spectra were analyzed using custom-built IDL routines. To correct the spectra for the sky background, we selected two regions 
well above and below the stellar spectrum in the spatial direction. For each spatial pixel, we fitted a linear trend to the flux values in these regions, 
determined the sky below the stellar spectrum via interpolation, and subtracted it. This procedure serves to remove any background trends of
instrumental or physical origin, however for the data described here, no appreciable trends were seen across the sky background in the spatial direction.
After sky removal, the spectra were extracted by summing flux inside large (18 px) apertures.

Visual inspection of the derived spectra quickly revealed residual wavelength offsets of up to 0.5~nm, between target 
and reference stars, and wavelength drifts up to 0.3~nm, affecting the spectra of individual stars throughout each observing sequence. As our smallest 
wavelength bin is 2~nm, these drifts are large enough to affect our results and we thus refined the wavelength solution on the target and 
reference spectra as follows before proceeding.
To calculate a correct wavelength solution for all images, we first removed telluric features with the molecfit routines \citep{Smette15,Kausch15}. 
We then calculated improved wavelength solutions based on these corrected spectra and applied these to the individual (uncorrected) spectra 
before extracting spectrophotometry. The improved wavelength solutions were found in a three-step process:
First, we identified the best frames of each sequence as reference images and, for each star, shifted all other spectra obtained during 
the same night to the wavelength solution of these reference spectra. We did this by cross-correlating a set of reference regions and, based on the 
inferred wavelength shifts, adapted the wavelength solution of each of the three detectors separately. 
Once all spectra of each sequence and star were aligned to the respective reference spectrum, we proceeded to finding an improved wavelength solution.
To do so, we first combined all spectra of each star and date, producing a higher S/N spectrum.
We then refined the wavelength solution of the target spectrum as follows. We fitted the continuum, normalized the target spectrum and cross-correlated a set of 
well-resolved features against a PHOENIX \citep{Husser13} model spectrum of an {\teff}=6100~K, {\logg}=4.50 solar-metallicity star that had been normalized 
and convolved down to the resolution of our data. Again, the wavelength solution was corrected by fitting cross-correlating a set of reference regions. 
Following the same procedure, we finally matched the two reference star spectra to that of the target.

\begin{figure}
 \includegraphics[width=\linewidth]{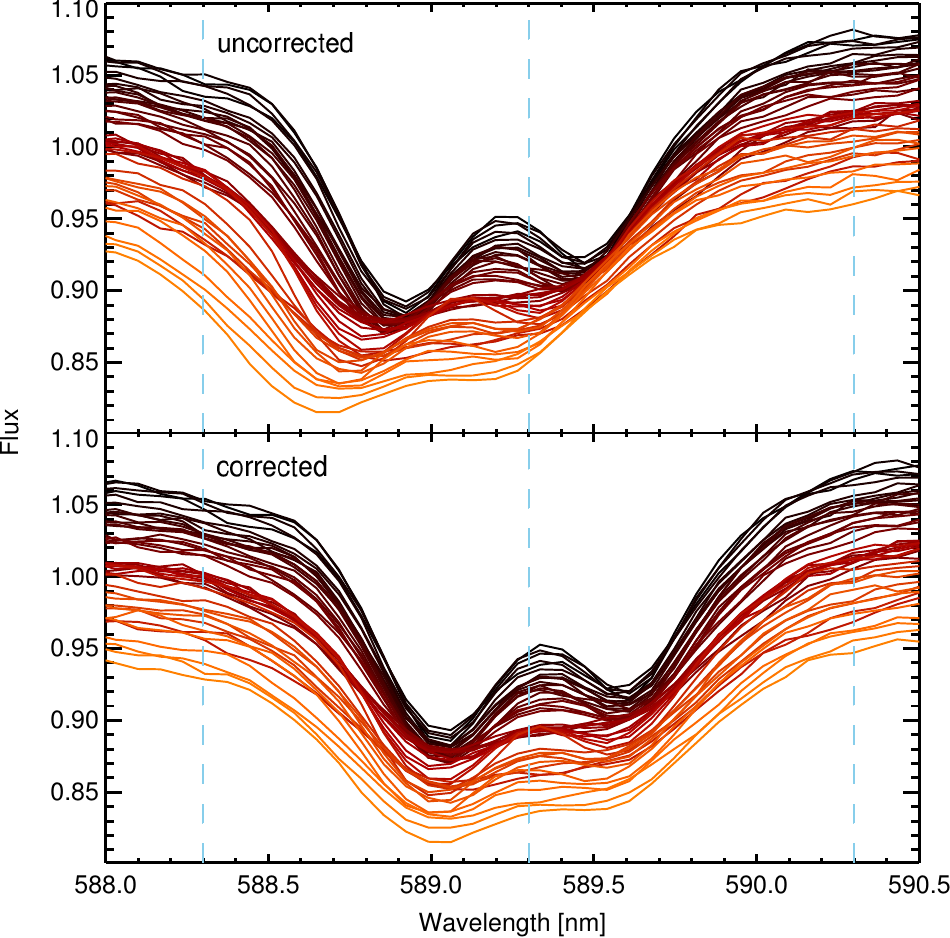}
 \caption{\label{fig:shifts}Enlargement of the Na feature at 589~nm showing all spectra obtained on 07 June 2015. Flux is given in arbitrary units,
 and the spectra are color-coded and shifted on the y-axis for illustrative purposes. The top panel shows the drift of the wavelength solution throughout 
 the sequence before wavelength recalibration, and the bottom panel shows the same sequence of spectra after recalibration.
 Vertical blue dashed lines indicate the center and limits of our smallest wavelength bin around the Na feature.}
\end{figure}

\begin{figure*}
 \includegraphics[width=\linewidth]{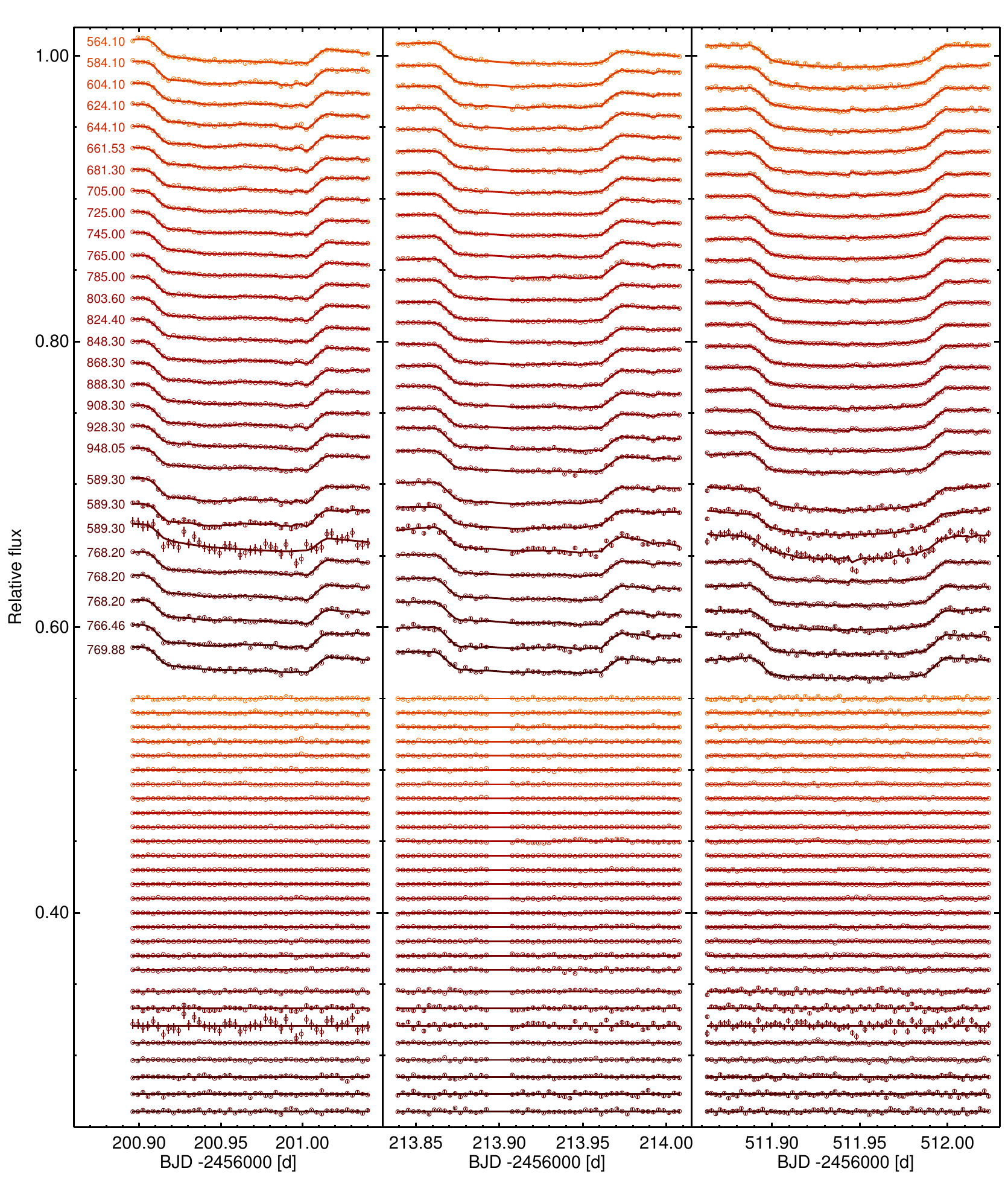}
 \caption{\label{fig:lcb}Spectrophotometric light curves of the three observation epochs (left, center, and right panels) obtained using 20~nm bins, 
 as well as bins centered on the Na and K line features. The central wavelength of each bin (in nm) is indicated in the left panel. The light curves centerend on the alkali features
 are ordered by bin width (from top to bottom): 10, 5 and 2~nm. Residuals are shown below, in the same order as the light curves.}
\end{figure*}

\begin{figure*}
 \includegraphics[width=\linewidth]{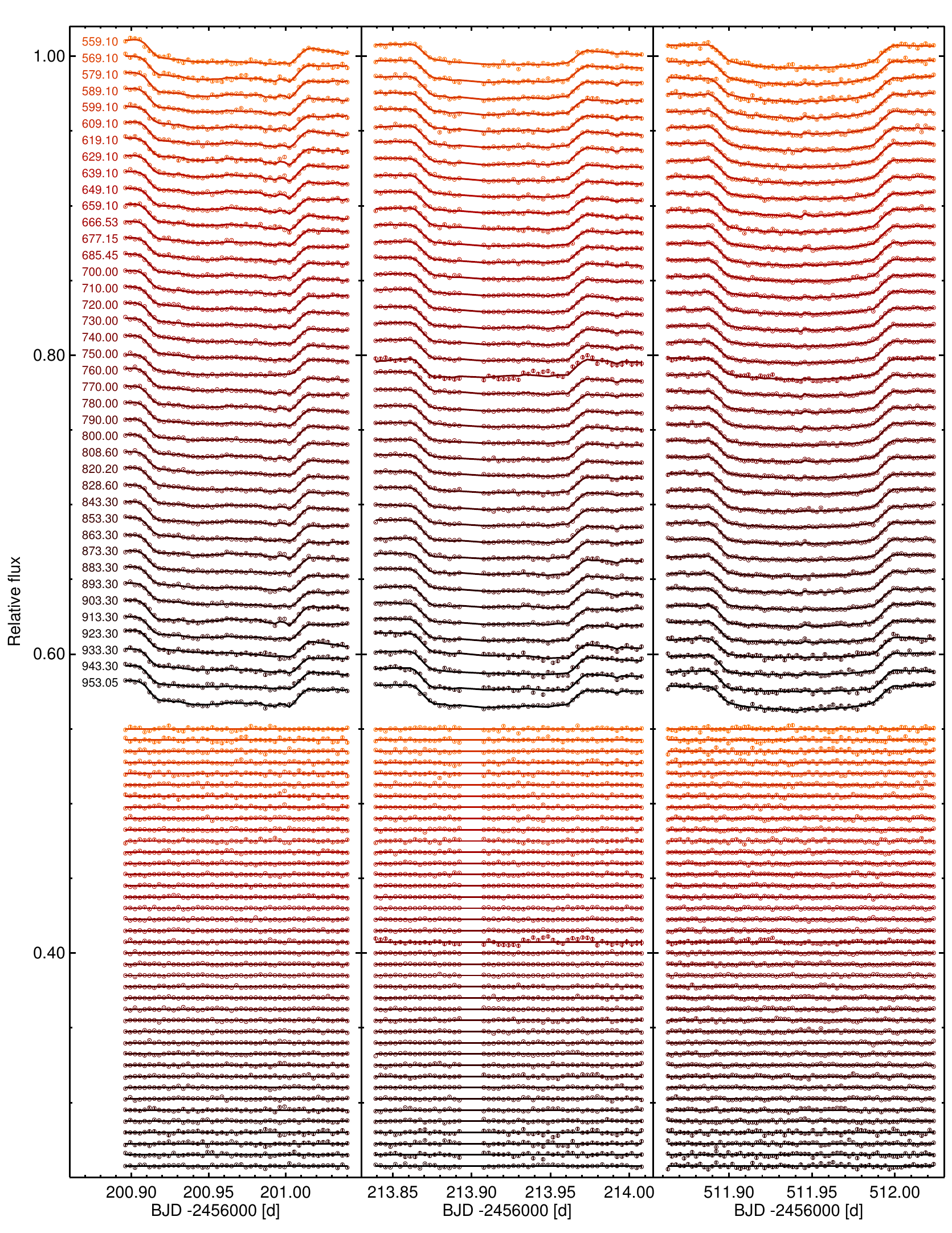}
 \caption{\label{fig:lca}Spectrophotometric light curves of the three observation epochs (left, center, and right panels) obtained using 10~nm bins. 
 The central wavelength of each bin (in nm) is indicated in the left panel. Residuals are shown below, in the same order as the light curves.}
\end{figure*}

Once we were certain of a reliable wavelength calibration, we binned the spectra of target and reference stars in 10 and 20~nm wide wavelength 
bins and created relative transit light curves for each bin. 
To reach a higher wavelength resolution on the Na and K features, we add 2, 5 and 10~nm bins centered on these features. We find that
the most precise light curves are obtained when only the fainter reference star is used instead of the the sum of both references, since the brighter reference
star showed strong correlated noise.
See Figs. \ref{fig:lcb} and \ref{fig:lca} for these spectrophotometric light curves, together with the best models described below.

\section{Lightcurve analysis}
\label{sec:mod}
\subsection{Fitting method}

Our code uses a differential evolution Markov chain Monte Carlo (DEMCMC) approach for the combined analysis of transit lightcurves,
combining the publicly available parallelized MC3 code \citep{Cubillos17}\footnote{https://github.com/pcubillos/MCcubed} with the prescription for transit lightcurve shape by \citet{Mandel02}. 
Depending on the available resources (differential evolution MCMC methods rely on a large number of chains run in parallel),
the sampling algorithm may be set to either a Metropolis-Hastings algorithm \citep[e.g.][]{Carlin09}, a differential-evolution MCMC (DEMCMC) \citep{terBraak2006}
algorithm, or the DEMC-z algorithm with snooker proposals \citep{terBraak08}. Convergence is checked via the Gelman-Rubin \citep{Gelman92} test.

To account for photometric variations stemming from instrumental, atmospheric or stellar effects, our code includes
photometric baseline models, parametrizations of external variables such as time, stellar FWHM, and sky background. 
The parametric baseline coefficients can either be included as jump parameters or calculated via least-square minimization at each MCMC step (following \citealt{Gillon10a}).
The later option can be advantageous for computational reasons as it reduces the number of jump parameters, however the first option allows to visualize 
any correlations between photometric baseline parameters and inferred transit parameters.
We assume a minimal baseline model assuming a second-order time polynomial, and accept more complicated models 
only if the Bayes factor \citep[e.g.,][]{Schwarz78} estimated from the Bayesian Information Criterion indicated significantly 
higher probability. It is also possible to include a common prescription of correlated noise for spectrophotometric light curves from 
the same epoch and instrument based on the common noise model (CNM) described in \citet{Lendl16}.
Before launching the MCMC, we run a least-squares minimization and estimate the amplitudes of additional white and 
red noise in each of the lightcurves via the $\beta_r$ and $\beta_w$ factors. $\beta_w$ is defined as 
the ratio of the residual RMS to the mean error and $\beta_r$ is calculated by comparing the residual RMS 
of the binned to the residuals of the unbinned data set \citep{Winn08,Gillon10a}. 
For each lightcurve, we then rescale the errors by multiplying them with the product 
$CF = \beta_w \times \beta_w$. This procedure also assures the correct weighting between several data sets of unequal quality.

The following parameters can be set as fitted (''jump'') parameters in the analysis:
\begin{itemize}
 \item the star to planet radius ratio $\frac{R_{p}}{R_{\ast}}$
 \item the impact parameter, $b$
 \item the transit duration, $T_{14}$
 \item the time of mid-transit, $T_0$
 \item the orbital period, $P$
 \item $\sqrt{e}\sin{\omega}$ and $\sqrt{e}\cos{\omega}$, where $e$ is the orbital eccentricity and $\omega$ is the argument of periastron
 \item the linear combination of the quadratic limb-darkening coefficients $(u_1,u_2)$ in each 
       wavelength band, $c_{1,i}=2u_{1,i} + u_{2,i}$ and $c_{2,i} = u_{1,i} - 2u_{2,i}$ \citep{Holman06}
 \item offsets ${ \left( \Delta \frac{R_{p}}{R_{\ast}}\right)}_{i}$ from a reference radius ratio 
       ${ \left( \frac{R_{p}}{R_{\ast}} \right)}_{0}$ for each wavelength channel. 
       ${ \left( \frac{R_{p}}{R_{\ast}} \right)}_{0}$ is usually determined from broadband photometry or 
       ``white''-light analysis binning all spectrophotometric data together. 
       It is not possible to fit ${ \left( \Delta \frac{R_{p}}{R_{\ast}}\right)}_{i}$ and $\frac{R_{p}}{R_{\ast}}$ simultaneously.
 \item the coefficients of the photometric baseline models
\end{itemize}

\subsection{WASP-103 analysis}

For the analysis of the GMOS WASP-103 data, we used the DEMCMC algorithm running between 30 and 180 parallel (depending on the number of 
fitted light curves) chains of 20000 points each.
To calculate limb darkening coefficients for all our wavelength channels, we used the routines by \citet{Espinoza15} together with the response 
function for the GMOS R400 grism\footnote{available from https://www.gemini.edu/sciops/instruments/gmos/spectroscopy-overview/gratings}. 
We calculated coefficients for a $\teff=6110.0$~K, M/H = 0.00 
and $v_{turb}=2.0$~{\kms} star for surface gravities of $\logg = 4.0$ and $\logg = 4.5$ using interpolated PHOENIX models (option P100).
We then interpolated these values to match the surface gravity of WASP-103, $\logg = 4.22$. We kept the limb-darkening values fixed during
our analysis. We verified that this did not impact our results by performing an additional run while letting the limb darkening parameters 
vary, assuming only a wide (width 0.1, much larger than the variation of the parameters between neighboring wavelength bins) 
Gaussian prior centered on the model values. Results from this test were near-identical to those of with fixed limb-darkening parameters.

\subsubsection{Contamination}
\label{sec:conta}

WASP-103 was identified to possess a nearby (0.24'') star by \citet{Woellert15}, which is blended with WASP-103 in our data. 
Further observations by \citet{Ngo2016} and \citet{Cartier17} reveal this contaminant to be a $T=4400\pm200$~K K5V star, most likely
physically associated to WASP-103 \citep{Cartier17}.
Third light, if unaccounted for, introduces systematic effects in transmission spectra which may be misinterpreted as effects 
stemming from the planetary atmosphere. To correct our measurements, we estimate the contaminating flux in 
the observed wavelength bands using PHOENIX model spectra \citep{Husser13} and the target and contaminant properties
of \citet{Cartier17}. We binned PHOENIX model spectra to the spectral bins of our transmission spectrum and interpolated them to the 
target and contaminant parameters. We then calculated the contaminating flux ratio for each wavelength bin $i$ as
\begin{equation}
 \left(\frac{F_{\mathit{cont}}}{F_{\mathit{W103}}}\right)_i = \left(\frac{R_{\mathit{cont}}}{R_{\mathit{W103}}}\right)^2 \left(\frac{M_{\mathit{cont}}}{M_{\mathit{W103}}}\right)_i \, ,
\end{equation}
where $(M_{\mathit{W103}}$,$M_{\mathit{cont}})_i$ are the binned model spectra, and $R_{\mathit{cont}}/R_{\mathit{W103}}$ is the 
contaminant/target radius ratio. To estimate the uncertainties, we drew 10000 temperature values from Gaussian distributions of target and contaminant 
values centered on the values given by \citet{Cartier17}, and with a standard deviation of their respective $1-\sigma$ errors. 
As the radius ratio is not independent of the stellar temperature, it needs to be estimated for each combination of target and contaminant temperatures.
We did this by interpolating and binning model spectra for all temperature values and using these together with the measured 
J, H and K band magnitude ratios of \citep{Cartier17} to infer appropriate radius ratios for our sample of target and contaminant temperatures.
For this set of 10000 simulated pairs of spectra and radius ratios, we inferred the resulting 
flux ratios in the spectral bins of our analysis and estimated the errors on the assumed flux ratios from this distribution.
The obtained values and uncertainties are given in Table \ref{tab:conta}.
These values are in good agreement with those given by \citet{Southworth16} and \citet{Cartier17}.
We note that errors in the assumed contamination do not affect each spectral measurement independently, but rather cause horizontal shifts 
of the full transmission spectrum, and/or introduce a wavelength-dependent slope. 
To evaluate any such effect on our transmission spectrum, we derive the impact of each element of our sample on the measured $R_P/R_\ast$ values and 
then infer the 3-$\sigma$ contours. These contours are shown as dotted lines in Fig. \ref{fig:traspec_master},
and indicate that, while the measurements may be shifted up or down by 0.0013 in $R_p/R_\ast$, potentially introduced slopes are small.

\begin{table*}[h]
\centering   
\caption{\label{tab:conta}}Wavelength bins, contamination flux ratios ($F_{cont}/F_{W103}$) for WASP-103 and the nearby source, and
quadratic limb-darkening coefficients used in the analysis.
\begin{tabular}{cccccccc} \hline \hline
Wavlength range [nm] & $F_{cont}/F_{W103}$ & $u_1$ & $u_2$ & Wavelength range [nm] & $F_{cont}/F_{W103}$ & $u_1$ & $u_2$  \T\\
\hline
\multicolumn{4}{c}{10~nm bins}  \T & \multicolumn{4}{c}{20~nm bins}   \\
554.1 -- 564.1 & $ 0.047\pm{0.0099}$ & 0.499 & 0.208  & 554.1 -- 574.1   &   $ 0.049\pm{0.0098} $ & 0.493 &  0.209  \T \\
564.1 -- 574.1 & $ 0.051\pm{0.0098}$ & 0.488 & 0.210  & 574.1 -- 594.1   &   $ 0.051\pm{0.0099} $ & 0.472 &  0.212  \\
574.1 -- 584.1 & $ 0.053\pm{0.0097}$ & 0.477 & 0.211  & 594.1 -- 614.1   &   $ 0.056\pm{0.0097} $ & 0.450 &  0.217  \\
584.1 -- 594.1 & $ 0.049\pm{0.0103}$ & 0.466 & 0.213  & 614.1 -- 634.1   &   $ 0.056\pm{0.0100} $ & 0.432 &  0.219  \\
594.1 -- 604.1 & $ 0.055\pm{0.0099}$ & 0.459 & 0.212  & 634.1 -- 654.1   &   $ 0.060\pm{0.0097} $ & 0.399 &  0.232  \\
604.1 -- 614.1 & $ 0.056\pm{0.0095}$ & 0.449 & 0.214  & 654.1 -- 668.9   &   $ 0.065\pm{0.0094} $ & 0.383 &  0.235  \\
614.1 -- 624.1 & $ 0.056\pm{0.0101}$ & 0.436 & 0.218  & 673.0 -- 689.6   &   $ 0.063\pm{0.0108} $ & 0.383 &  0.226  \\
624.1 -- 634.1 & $ 0.057\pm{0.0100}$ & 0.429 & 0.218  & 695.0 -- 715.0   &   $ 0.067\pm{0.0095} $ & 0.366 &  0.197  \\
634.1 -- 644.1 & $ 0.059\pm{0.0100}$ & 0.426 & 0.187  & 715.0 -- 735.0   &   $ 0.070\pm{0.0089} $ & 0.350 &  0.199  \\
644.1 -- 654.1 & $ 0.061\pm{0.0094}$ & 0.398 & 0.201  & 735.0 -- 755.0   &   $ 0.073\pm{0.0080} $ & 0.350 &  0.213  \\
654.1 -- 664.1 & $ 0.065\pm{0.0093}$ & 0.384 & 0.236  & 755.0 -- 775.0   &   $ 0.075\pm{0.0082} $ & 0.339 &  0.214  \\
664.1 -- 668.9 & $ 0.064\pm{0.0096}$ & 0.393 & 0.227  & 775.0 -- 795.0   &   $ 0.077\pm{0.0080} $ & 0.318 &  0.232  \\
673.0 -- 681.3 & $ 0.063\pm{0.0106}$ & 0.387 & 0.226  & 795.0 -- 812.2   &   $ 0.079\pm{0.0074} $ & 0.308 &  0.234  \\
681.3 -- 689.6 & $ 0.062\pm{0.0112}$ & 0.381 & 0.226  & 816.0 -- 832.8   &   $ 0.081\pm{0.0072} $ & 0.292 &  0.236  \\
695.0 -- 705.0 & $ 0.066\pm{0.0094}$ & 0.371 & 0.227  & 838.3 -- 858.3   &   $ 0.083\pm{0.0072} $ & 0.276 &  0.242  \\
705.0 -- 715.0 & $ 0.067\pm{0.0097}$ & 0.361 & 0.198  & 858.3 -- 878.3   &   $ 0.086\pm{0.0070} $ & 0.278 &  0.237  \\
715.0 -- 725.0 & $ 0.069\pm{0.0094}$ & 0.354 & 0.198  & 878.3 -- 898.3   &   $ 0.087\pm{0.0070} $ & 0.279 &  0.232  \\
725.0 -- 735.0 & $ 0.071\pm{0.0085}$ & 0.350 & 0.230  & 898.3 -- 918.3   &   $ 0.089\pm{0.0067} $ & 0.275 &  0.231  \\
735.0 -- 745.0 & $ 0.072\pm{0.0081}$ & 0.343 & 0.230  & 918.3 -- 938.3   &   $ 0.091\pm{0.0062} $ & 0.273 &  0.228  \\
745.0 -- 755.0 & $ 0.073\pm{0.0080}$ & 0.347 & 0.213  & 938.3 -- 957.8   &   $ 0.093\pm{0.0061} $ & 0.261 &  0.232  \\
755.0 -- 765.0 & $ 0.075\pm{0.0078}$ & 0.348 & 0.206  &                &                       & &   \\                                   
765.0 -- 775.0 & $ 0.075\pm{0.0084}$ & 0.331 & 0.225  & \multicolumn{4}{c}{Na feature}          \\
775.0 -- 785.0 & $ 0.077\pm{0.0082}$ & 0.320 & 0.232  & 584.3 -- 594.3    &  $ 0.049\pm{0.010} $ & 0.465 & 0.215  \T \\
785.0 -- 795.0 & $ 0.078\pm{0.0077}$ & 0.314 & 0.233  & 586.8 -- 591.8    &  $ 0.044\pm{0.011} $ & 0.479 & 0.200 \\
795.0 -- 805.0 & $ 0.079\pm{0.0075}$ & 0.314 & 0.224  & 588.3 -- 590.3    &  $ 0.032\pm{0.010} $ & 0.466 & 0.214 \\
805.0 -- 812.2 & $ 0.079\pm{0.0073}$ & 0.310 & 0.225  &                 &                       & &  \\
816.0 -- 824.4 & $ 0.080\pm{0.0073}$ & 0.296 & 0.236  &  \multicolumn{4}{c}{K feature}          \\
824.4 -- 832.8 & $ 0.082\pm{0.0072}$ & 0.288 & 0.239  & 763.2 -- 773.2     & $ 0.075\pm{0.0084} $ & 0.328 & 0.230 \T \\
838.3 -- 848.3 & $ 0.083\pm{0.0073}$ & 0.281 & 0.234  & 765.7 -- 770.7     & $ 0.074\pm{0.0085} $ & 0.329 & 0.220 \\
848.3 -- 858.3 & $ 0.084\pm{0.0072}$ & 0.280 & 0.234  & 767.2 -- 769.2     & $ 0.076\pm{0.0084} $ & 0.325 & 0.231 \\
858.3 -- 868.3 & $ 0.086\pm{0.0070}$ & 0.270 & 0.241  & 765.46 -- 767.46 & $ 0.072\pm{0.0084} $ & 0.327 & 0.232 \\
868.3 -- 878.3 & $ 0.086\pm{0.0071}$ & 0.273 & 0.236  & 768.88 -- 770.8.8 & $ 0.074\pm{0.0086} $ & 0.326 & 0.232 \\
878.3 -- 888.3 & $ 0.087\pm{0.0069}$ & 0.268 & 0.239  &                  &                     & &  \\
888.3 -- 898.3 & $ 0.088\pm{0.0070}$ & 0.265 & 0.240  &  \multicolumn{4}{c}{white}              \\
898.3 -- 908.3 & $ 0.090\pm{0.0069}$ & 0.280 & 0.225  & 5541 -- 9578     &  $ 0.069\pm{0.0087} $ & 0.341 & 0.226  \T \\
908.3 -- 918.3 & $ 0.089\pm{0.0066}$ & 0.279 & 0.223  &                  &                      & & \\
918.3 -- 928.3 & $ 0.092\pm{0.0062}$ & 0.275 & 0.225  &                  &                      & & \\
928.3 -- 938.3 & $ 0.091\pm{0.0064}$ & 0.273 & 0.223  &                  &                      & & \\
938.3 -- 948.3 & $ 0.092\pm{0.0061}$ & 0.265 & 0.207  &                  &                      & & \\
948.3 -- 957.8 & $ 0.095\pm{0.0061}$ & 0.254 & 0.212  &                  &                      & & \\
\hline                                                  
\end{tabular}                                            
\end{table*}                                              

\subsubsection{White light curves}
\label{sec:whites}
We first performed an independent analysis of each of the white light curves obtained by combining the full spectra of each date.
Here, we tested a wide variety of parametric baseline models to find the optimal fit to the observations of each date, selecting 
a more complicated model over a simple one only if the Bayes factor indicates significantly higher probability. Error bars
were not adapted to compensate for red or white noise at this step, as this would bias the inferred BIC values.
The observations of 27 June 2015 and 10 July 2015 are best fit by polynomials of second order in time together with a sine 
function, $f(t)=A_0+A_1 t + A_2 t^2 + B_0 \sin(B_1 t + B_2)$ (a common behavior for GMOS data, see \citealt{Stevenson14,vonEssen17}), 
while for the light curve of 03 May 2016, a second order time polynomial is adequate. Using these baseline models, and allowing for 
error scaling to compensate for excess noise, we infer $R_p/R_\ast$ values of $0.1120 \pm 0.0021$, $0.1158 \pm 0.0011$, and $0.1157 \pm 0.0007$ for the light curves 
observed on 27 June 2015, 10 July 2015, and 03 May 2016, respectively. These values are in good agreement, with the two 
latter light curves producing near-identical values and the first light curve, which is also the least precise, showing a marginally 
smaller value. This is in line with the non-detection of rotational modulation (which would lead to time-variable transit depths) for 
the host star \citep{Gillon14}. From a combined fit of all three white light curves, we derive the transit parameters listed in 
Table \ref{tab:whites}. Again, these values are in good agreement ($<1.5 \sigma$ difference) with those published by 
\citet{Southworth15,Southworth16} and \citet{Cartier17}.

\begin{table}[h]
\centering   
\caption{\label{tab:whites}}The \mbox{transit} parameters inferred from an analysis of the white light curves. 
\begin{tabular}{ll} \hline \hline
Parameter & Value \T \\
\hline
Time of mid-transit, $T_0$ [$\mathrm{HJD_{UTC}}$] & $ 7511.943617 \pm{0.000076} $ \T \\
Star/planet radius ratio, $R_P/R_\ast$    & $   0.11442_{-0.00043}^{+0.00050}   $ \\
Impact parameter, $b$                     & $   0.074_{-0.048}^{+0.066}         $ \\
Transit duration, $T_{\mathit{14}}$ [d]   & $   0.11232\pm{0.00019}             $ \\
Period [d]                                & $   0.92554705 \pm{0.0000004}       $ \\
\hline                                                  
\end{tabular}                                            
\end{table}        

\subsubsection{Inference of the transmission spectrum}

We extracted spectrophotometry covering the full spectral range for wavelength intervals of 10~nm and 20~nm width, as well as for 
2~nm, 5~nm and 10~nm windows centered on the Na feature. Due to the separation of the K feature components (which is 3.4~nm, compared to 
0.6~nm for the Na feature), we placed two 2~nm bins at the center of the two components and one bin between them, in addition to 5~nm and 10~nm
bins covering the whole feature. We analyzed the full-range 10 and 20-nm bins as separate sets, each time including the narrow bins
centered on Na and K. Instead of fixing the transit shape parameters ($T_0$, $b$, and $T_{\mathit{14}}$) when inferring the transmission spectrum, which may result in under-estimated errors, 
we let these parameters vary, but imposed normal priors centered on the inferred value and 1-$\sigma$ errors obtained from the analysis of the white light curves.
We used the white $R_p / R_\ast$ values measured above as reference radii used in the calculation of the common noise model and the planetary radius variations 
$(\Delta R_p/R_\ast)_i$. 

\begin{figure}
 \includegraphics[width=\linewidth]{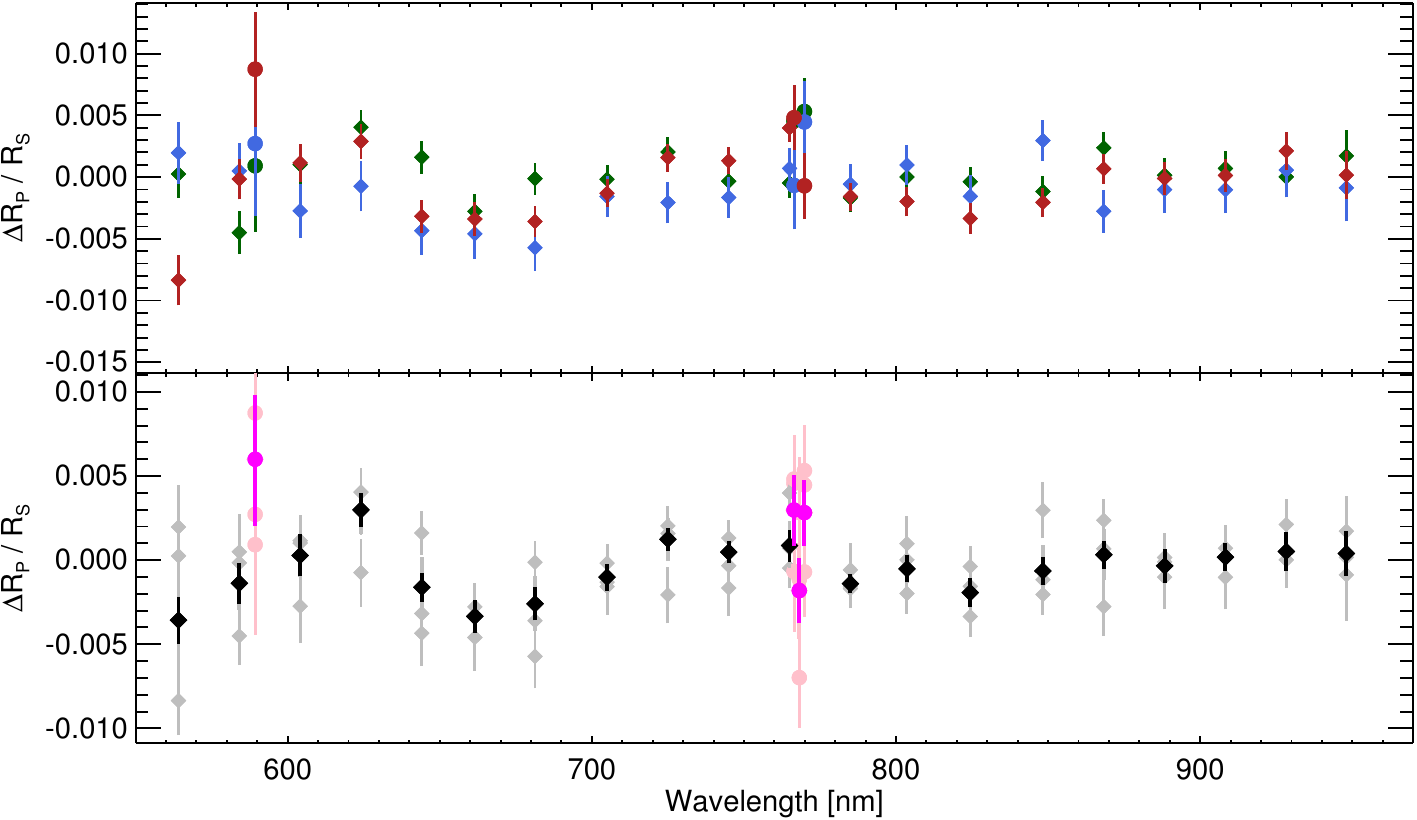}
 \caption{\label{fig:traspec}Inferred $R_p/R_\ast$ offsets for wavelength bins of using 20~nm width. Top: results from
 separate analyses to each epoch, where blue, red and green points refer to results from 27 June 2015, 10 July 2015, and 03 May 2016, respectively.
 Bottom: results from the combined analysis of all data (black points), together with the 2~nm bins centered on the Na and K 
 features. Results from the separate analyses are shown in gray. }
\end{figure}

We first carried out individual analyses of the transmission spectra obtained at the individual epochs, to search for any variations.
For each set, we used the best parametric model found from the analysis of the white light curves (see Sect. \ref{sec:whites}),
fitting all coefficients of the time polynomial as well as the amplitude of the sine function. We also tested the addition of a common noise model, finding that 
its addition is warranted by significant BIC improvements for all three dates.
Additionally, the resulting transmission spectra obtained when including a common noise model show better agreement between the three epochs.
We find that the spectra obtained on the three dates are in reasonable overall agreement with one another (see the upper panel of Fig. \ref{fig:traspec}), 
with the discrepancy of two measurements in the same wavelength bin at a level of 0.65 $\sigma$ and below 2.3 $\sigma$ in all cases. For the 10~nm
bins, agreement is at a similar level, with a median of 0.55~$\sigma$ and individual measurements disagreeing by as much as 2.5~$\sigma$.
The differences are most pronounced at short or long wavelengths, which could indicate that limited signal to noise combined with 
larger supplementary wavelength corrections as limiting factors.

\begin{figure}
 \includegraphics[width=\linewidth]{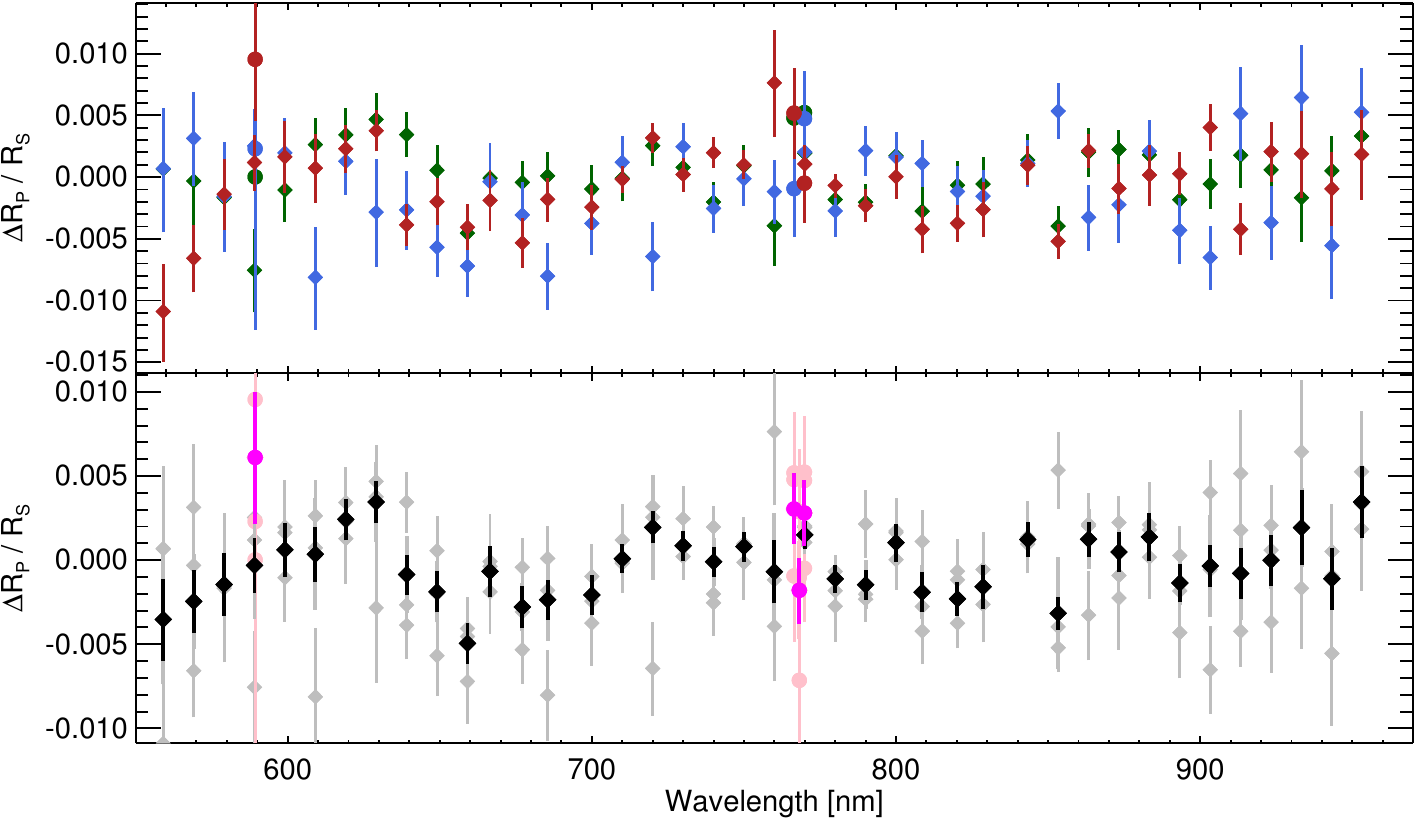}
 \caption{\label{fig:traspec_narr}Inferred $R_p/R_\ast$ offsets for wavelength bins of using 20~nm width. Top: results from
 separate analyses to each epoch, where blue, red and green points refer to results from 27 June 2015, 10 July 2015, and 03 May 2016, respectively.
 Bottom: results from the combined analysis of all data (black points), together with the 2~nm bins centered on the Na and K 
 features. Results from the separate analyses are shown in gray. }
\end{figure}

To find the optimal transmission spectrum based on all available data, we performed a combined analysis of all GMOS data, allowing for a single planetary 
radius offset $(\Delta R_p/R_\ast)_i$ per wavelength channel. The resulting transmission spectra are shown in Figs. \ref{fig:traspec} (20~nm bins) and 
\ref{fig:traspec_narr} (10~nm bins), and given in Table \ref{tab:traspec}.

\section{Results and Discussion}
\label{sec:res}

\subsection{Model atmospheres}
We compute model transmission spectra of WASP-103b, using solar abundances, to aid the interpretation of our observations.
To produce atmospheric transmission forward models of WASP-103b, we use the open-source 
Python Radiative Transfer in a Bayesian framework (Pyrat Bay) package\footnote{\href{http://pcubillos.github.io/pyratbay}
{http://pcubillos.github.io/pyratbay}} (Cubillos et al. 2017, in prep.), which is based on the 
Bayesian Atmospheric Radiative Transfer package \citep{Blecic16a, Cubillos16}. The models of WASP-103b 
consider molecular opacities for H$_{\rm 2}$O \citep[HITEMP,][]{Rothman10}, collision induced
absorption from H$_{\rm 2}$-H$_{\rm 2}$ \citep{Borysow01, Borysow02} and H$_{\rm 2}$-He \citep{Borysow88,
Borysow89a, Borysow89b}, H$_{\rm 2}$ Rayleigh scattering \citep{Lecavelier08b}, and resonant
alkali lines \citep{Burrows00}.  

The code computes line-by-line radiative-transfer for a 1D atmospheric model consisting of concentric 
shell layers, in hydrostatic equilibrium, and thermochemical-equilibrium abundances \citep{Blecic16b}.
We assumed isothermal temperature profiles with a temperature of 2500
K, and calculated thermochemical-equilibrium models from solar
elemental abundances scaling the metallicities to 0.1, 1, and 10 times
the solar value.

\subsection{Spectral slope by scattering or contamination}
\begin{figure*}
 \centering
 \includegraphics[width=\linewidth]{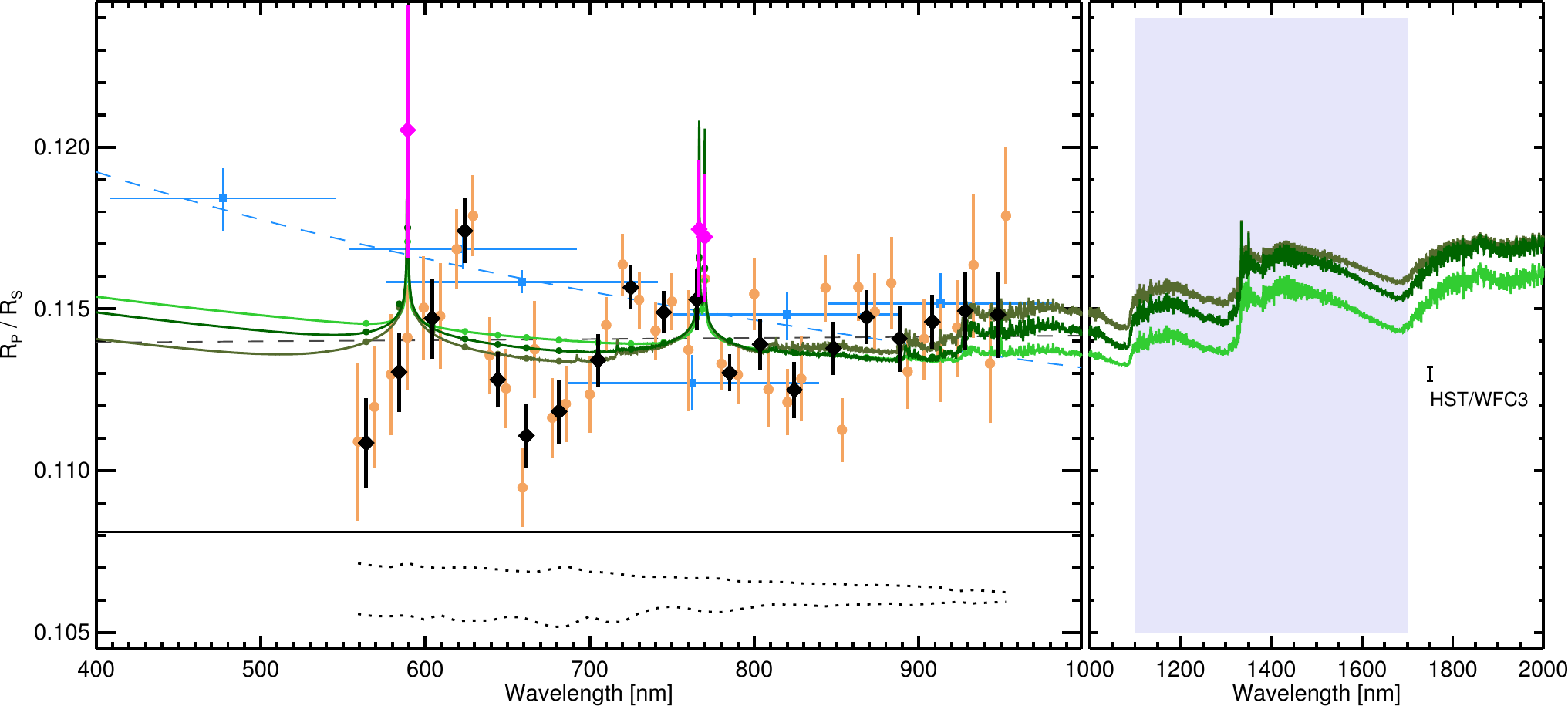}
 \caption{\label{fig:traspec_master}Optical transmission spectrum of WASP-103b. Measurements obtained with 10 and 20~nm wide 
 bins are shown as brown filled circles and black squares, respectively. Measurements in 2~nm wide bins centered on the Na feature,
 as well as on both components of the K feature are shown in magenta. Observations by \citet{Southworth16} are 
 indicated by blue squares, together with their inferred spectral sloped (blue dashed line). The best-fit spectral slope inferred from
 our data is shown as a gray dashed line. Model transmision spectra for WASP-103b's are shown as a light green (0.1 $\times$ solar metallicity),
 dark green (solar metallicity) and olive (10 $\times$ solar metallicity) lines and filled circles indicate the models binned to 
 the resolution of the data. Below, the two dotted lines indicate the possible added spectral slopes (at 3-$\sigma$ level) due to uncertainties in the 
 properties of WASP-103b and the contaminating star. The vertical scale is identical to that of the upper panel.
 In the right panel, the shaded region indicates the wavelength coverage of HST/WFC3's G141 grism,
 and the expected precision of spectrophotometric WFC3 measurements is indicated.}
\end{figure*}

We use our combined transmission spectra to study the atmospheric properties of WASP-103b. 
\citet{Southworth15} used broadband light curves to infer the presence of a strong
scattering slope in the transmission spectrum of WASP-103b, which holds even after correcting
their measurements from contamination by the nearby star \citep{Southworth16}. While our data 
do not span the full extent in wavelength of their measurements, in particular we
have no data bluewards of 560~nm, we do not observe any evident trend in the transmission spectrum.
To test this, we fitted a straight line with a linear model in wavelength to the 10~nm bin data set, 
excluding the narrow wavlength bins centered on the Na and K features. We found a negligible slope of absorption 
increasing with wavelength $1.04 \times 10^{-6} \pm 1.54 \times 10^{-6} $ in $R_P/R_\ast$ per nm
(best values and errors were determined via an MCMC bootstrap approach drawing 10000 subsamples). 
Comparing this solution with a straight horizontal line, we find no evidence of the slope producing a 
significantly more accurate fit to the data (Bayes factor 0.2). 
We show our measurements next to those of \citet{Southworth16} in Fig. \ref{fig:traspec_master}. Next to the data, we show
our best-fit slope as well as a Rayleigh scattering slope inferred from the broadband data as in \citet{Southworth16}.
The latter is parametrized, following \citet{Lecavelier08a}, by $\frac{dR_p/R_\ast(\lambda)}{d \ln \lambda} = -0.0066(17)$.
Additional observations at shorter wavelengths would be helpful to definitely 
measure any scattering signature in the atmosphere of WASP-103b.

Beside the physical properties of the planetary atmosphere, blended third light from a companion star affects the transmission spectrum of
WASP-103b. While we correct for this third light in our analysis, uncertainties in the assumed spectral type of 
target and contaminating star can lead to residual slopes and overall offsets. As described in 
Sect. \ref{sec:conta}, we have carried out a large set of simulations, perturbing the assumed 
temperatures of target and contaminant within the errors given by \citet{Cartier17}. We find that
even in the most extreme cases, the added slopes are too small to significantly impact the shape of our observed
transmission spectrum. However, our transmission spectrum may be subject to an overall offset of $\pm0.0013$ in $R_P/R_\ast$.
This is illustrated in Fig. \ref{fig:traspec_master}, where two dotted gray lines indicate the potentially added slopes (assuming
the 3-$\sigma$ envelop to all simulated cases).

\subsection{Na and K absorption features}

\begin{figure}[h!]
 \includegraphics[width=\linewidth]{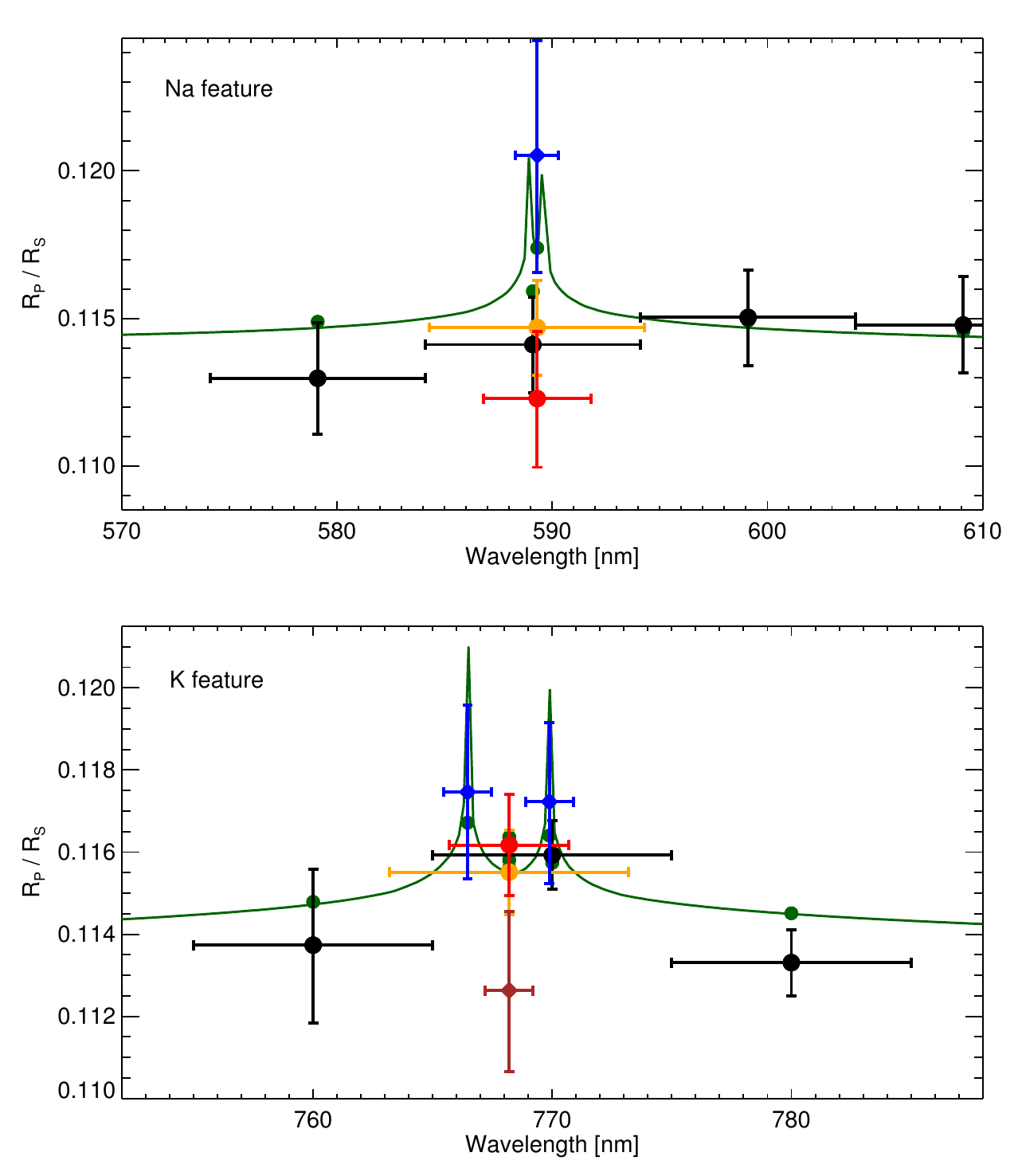}
 \caption{\label{fig:traspec_NaK} Transmission spectrum of WASP-103b zoomed in on the Na (top) and K (bottom) features. The 
 black points refer to the 10~nm wavelength bins also shown in Fig. \ref{fig:traspec_narr}, blue points indicate 2~nm
 bins centered on the features: either one bin centered on the Na feature, or two point centered on each component of the K feature. In the lower
 panel, brown refers to a 2~nm wide bin centered between the two components. Red and orange points denote wavelength bins
 of 5 and 10~nm width, respectively, centered on the features. The solar-composition model spectrum is shown in green, and filled circles indicate 
 the model binned to the resolution of the data.}
\end{figure}

Absorption by Na and K produces prominent features in planetary transmission spectra. As WASP-103b has a high mass and hence density, 
the predicted amplitude of these features is small compared
to other planets studied previously with ground-based transmission spectroscopy (e.g., 2.7 times smaller than those of the recently-studied 
WASP-39b, \citealp{Nikolov16}, and WASP-17b ,\citealp{Sedaghati16}), these features are inherently difficult to detect for WASP-103b. Our data 
cover both the 590~nm Na and the 766~nm K doublets, and we place a set of narrow bins on these features to test for extra absorption.
As shown in Fig. \ref{fig:traspec_NaK}, we detect signs of added absorption at the line cores of both features. 
The Na feature is more clearly detected. Even though the broadband transmission spectrum shows a trend (of unknown origin) across 
the 560 -- 630~nm region, the measured absorption is enhanced for the 2~nm bin centered on the Na feature compared to
the neighboring points (see Fig. \ref{fig:traspec_NaK}). 
We resolve the two components of the K feature by placing a 2~nm wide bin on each of the two peaks as 
well as on 2~nm wide bin between them. Further, we place bins of 5 and 10~nm centrally on the feature.
Both components show enhanced absorption compared to the neighboring spectral bins, or the spectral bin placed between the two components.

For both features, the absorption appears to be slightly larger (0.74, 0.36 and 0.43 $\sigma$ for the Na and the blue and red components 
of the K line, respectively) than predicted from our model spectra.
Enhanced absorption features could be related to the planet's tidally distorted state, for which we calculate 
(using the Roche lobe model by \citealp{Budaj11}, as well as stellar parameters from \citealp{Southworth16}) a substellar 
radius 10.5\% larger than the planet's polar radius. More detailed simulations of the effect of tidal distortion on 
transmission spectra are however beyond the scope of this work. 

As demonstrated by \citet{Deming17}, observations at low to medium spectral resolution are affected by an resolution linked bias (RLB),
which acts to decrease the apparent absorption feature amplitude. This is due to flux leakage from neighboring spectral regions into the absorption line,
where the planetary atmosphere's enhanced absorption is located. We have evaluated the RLB affecting our measurements of the Na and K lines following the formalism of \citet{Deming17} 
(Equations 3 -- 5), and assuming an instrumental spectral resolution of 1200 together with our smallest bin size of 2~nm. We find that the 
effect is small compared to our uncertainties, $2.0 \times 10^{-4}$, $5.4 \times 10^{-5}$, and $2.9\times 10^{-5}$ in $\Delta R_p/R_\ast$   
for the 2~nm~bins centered on the Na and the blue and red components of the K feature, respectively.

\subsection{Atmospheric metallicity and cloud altitude}
We compare our data to model atmospheres calculated for metallicities 0.1, 1 and 10 times the solar value. The data are most closely
reproduced by models with super-solar metal abundances, evidenced by $\chi^2$ values of 51.1, 56.0 and 60.0 when comparing data to 
models of 10, 1, and 0.1 solar metallicity. This is due to slightly enhanced absorption measured at wavelengths larger than
900~nm, where water absorption becomes prominent. Extending WASP-103's transmission spectrum to longer wavelengths, targeting
prominent water absorption features, would be required to make a more solid statement on the planet's atmospheric composition. 
We show the predicted transmission spectra in the near-IR in the right panel of Fig. \ref{fig:traspec_master}, indicating the location 
of HST's WFC3 G141 grism, and precision estimate of 50~ppm (e.g., \citealt{Kreidberg15}). Water should be easily detectable in the atmosphere
of WASP-103b.

We do not observe any evidence for aerosols in the planetary transmission spectrum. This means that, if clouds or hazes are present in the
atmosphere of WASP-103b, they must be located at altitudes below (or pressure levels larger than) those probed by our measurements.
For the models calculated here, the atmospheric pressure levels probed outside the alkali features
range between 0.01 bar (for 10 $\times$ solar metallicity) and 0.1 bar (for 0.1 $\times$ solar metallicity). If a cloud deck is
present in the atmosphere of WASP-103b, it must reside at pressures higher than 0.01 bar. This upper limit is well inside the range
of inferred cloud deck altitudes for hot Jupiters \citep[e.g.,][]{Barstow17}.

\subsection{The atmosphere of WASP-103b in context}

Based on these measurements, we tentatively classify WASP-103b as a planet with a clear atmosphere. 
The absence of a wavelength-dependent slope in our data, combined with the detection of alkali absorption indicates that aerosols, if present in substantial
quantities in the atmosphere of WASP-103b, exist largely at altitudes below those probed by tranmsission spectroscopy. 
The planet follows the trends of \citet{Stevenson16}, who proposed that hot planets, and planets with high surface gravities, are more 
likely to possess large water absorption features expected from clear atmospheres.
Along the same lines, the planet also follows the cloudiness-temperature relation proposed by \citet{Heng16}, who find that alkali
absorption features are most pronounced for hot planets. 

In terms of mass and temperature, WASP-103b greatly resembles WASP-12b, an object for which a detailed transmission spectrum has been observed
\citep{Mandell13,Sing13,Swain13,Stevenson14,Kreidberg15}. At the precision of the available data, the optical transmission spectrum of WASP-12b does not 
show alkali absorption features, however it does show a wavelength-dependent slope, which can be interpreted 
as signs of an aerosol layer \citep{Sing13,Barstow17}. Water has been detected \citep{Kreidberg15} in near-IR observations, but also 
these observations are best fit when including an additional atmospheric opacity source. It appears thus that the spectrum of WASP-12b is more
strongly affected by clouds or hazes than that of WASP-103b. However, direct comparison between results for the two planets is not straight-forward as our observations of 
WASP-103b have a different resolution and wavelength coverage than those by \citep{Sing13} of WASP-12b. Our data have a higher spectral resolution, 
and we are thus more sensitive to absorption in the alkali line cores, which may have gone undetected for WASP-12b. At the same time, our data 
cover a more narrow wavelength-region and are thus less sensitive to scattering slopes. Based on our detection of alkali features in the optical part 
of the spectrum, we would expect WASP-103b to show also a prominent water feature in its near-IR transmission spectrum.
Near-IR observations, but also observations spanning a wider wavelength region in the optical will allow a more detailed 
characterization of WASP-103b's atmosphere. Also, higher-resolution optical observations of WASP-12b would be an asset to make the available data  
on these two planets truly comparable.

\begin{table*}[h]
\centering   
\caption{\label{tab:traspec}}The transmission spectrum of WASP-103b as observed with GMOS.
\begin{tabular}{cccc} \hline \hline
Wavlength range [$\AA$] & $(\Delta R_p/R_\ast)_i$  & Wavelength range [$\AA$] & $(\Delta R_p/R_\ast)_i$ \T\\
\hline
\multicolumn{2}{c}{10~nm bins}  \T & \multicolumn{2}{c}{20~nm bins}   \\
5541 -- 5641 & $-0.0035 \pm{0.0024}  $  & 5541 -- 5741   &   $-0.00356 \pm{-0.0014} $  \T \\
5641 -- 5741 & $-0.0024 \pm{0.0019}  $  & 5741 -- 5941   &   $-0.00137 \pm{-0.0012} $  \\
5741 -- 5841 & $-0.0014 \pm{0.0019}  $  & 5941 -- 6141   &   $ 0.00029 \pm{-0.0012} $  \\
5841 -- 5941 & $-0.0003 \pm{0.0016}  $  & 6141 -- 6341   &   $ 0.00299 \pm{-0.0010} $  \\
5941 -- 6041 & $ 0.0006 \pm{0.0016}  $  & 6341 -- 6541   &   $-0.00161 \pm{-0.00087} $  \\
6041 -- 6141 & $ 0.0004 \pm{0.0016}  $  & 6541 -- 6689   &   $-0.00334 \pm{-0.00097} $  \\
6141 -- 6241 & $ 0.0024 \pm{0.0012}  $  & 6730 -- 6896   &   $-0.00259 \pm{-0.00099} $  \\
6241 -- 6341 & $ 0.0035 \pm{0.0013}  $  & 6950 -- 7150   &   $-0.00101 \pm{-0.00080} $  \\
6341 -- 6441 & $-0.0009 \pm{0.0012}  $  & 7150 -- 7350   &   $ 0.00124 \pm{-0.00068} $  \\
6441 -- 6541 & $-0.0019 \pm{0.0012}  $  & 7350 -- 7550   &   $ 0.00047 \pm{-0.00066} $  \\
6541 -- 6641 & $-0.0049 \pm{0.0012}  $  & 7550 -- 7750   &   $ 0.00087 \pm{-0.00095} $  \\
6641 -- 6689 & $-0.0007 \pm{0.0015}  $  & 7750 -- 7950   &   $-0.00140 \pm{-0.00057} $  \\
6730 -- 6813 & $-0.0028 \pm{0.0012}  $  & 7950 -- 8122   &   $-0.00051 \pm{-0.00080} $  \\
6813 -- 6896 & $-0.0024 \pm{0.0012}  $  & 8160 -- 8328   &   $-0.00192 \pm{-0.00088} $  \\
6950 -- 7050 & $-0.0021 \pm{0.0012}  $  & 8383 -- 8583   &   $-0.00064 \pm{-0.00082} $  \\
7050 -- 7150 & $ 0.00008\pm{0.00086} $  & 8583 -- 8783   &   $ 0.00033 \pm{-0.00083} $  \\
7150 -- 7250 & $ 0.00195\pm{0.00095} $  & 8783 -- 8983   &   $-0.0003 \pm{-0.0010} $  \\
7250 -- 7350 & $ 0.00086\pm{0.00088} $  & 8983 -- 9183   &   $ 0.00018 \pm{-0.00083} $  \\
7350 -- 7450 & $-0.00009\pm{0.00090} $  & 9183 -- 9383   &   $ 0.0005 \pm{-0.0012}  $  \\
7450 -- 7550 & $ 0.00081\pm{0.00088} $  & 9383 -- 9578   &   $ 0.0004 \pm{-0.0013} $  \\
7550 -- 7650 & $-0.0007 \pm{0.0019}  $  &                &                         \\                                   
7650 -- 7750 & $ 0.00151\pm{0.00083} $  & \multicolumn{2}{c}{Na feature}           \\
7750 -- 7850 & $-0.00111\pm{0.00080} $  & 5843 -- 5943    &  $ 0.0003 \pm{-0.0016} $   \T \\
7850 -- 7950 & $-0.00146\pm{0.00089} $  & 5868 -- 5918    &  $-0.0021 \pm{-0.0023} $  \\
7950 -- 8050 & $ 0.0010 \pm{0.0011}  $  & 5883 -- 5903    &  $ 0.0061 \pm{-0.0039} $  \\
8050 -- 8122 & $-0.0019 \pm{0.0012}  $  &                 &                        \\
8160 -- 8244 & $-0.0023 \pm{0.0010}  $  &  \multicolumn{2}{c}{K feature}           \\
8244 -- 8328 & $-0.0016 \pm{0.0013}  $  & 7632 -- 7732     & $ 0.0011 \pm{-0.0010} $  \T \\
8383 -- 8483 & $ 0.0012 \pm{0.0010}  $  & 7657 -- 7707     & $ 0.0018 \pm{-0.0012} $  \\
8483 -- 8583 & $-0.00316\pm{0.00098} $  & 7672 -- 7692     & $-0.0018 \pm{-0.0020} $  \\
8583 -- 8683 & $ 0.0012 \pm{0.0010}  $  & 7654.6 -- 7674.6 & $ 0.0030 \pm{-0.0021} $  \\
8683 -- 8783 & $ 0.0005 \pm{0.0012}  $  & 7688.8 -- 7708.8 & $ 0.0028 \pm{-0.0020} $  \\
8783 -- 8883 & $ 0.0014 \pm{0.0014}  $  &                  &              \\
8883 -- 8983 & $-0.0014 \pm{0.0011}  $  &                  &              \\
8983 -- 9083 & $-0.0003 \pm{0.0012}  $  &                  &              \\
9083 -- 9183 & $-0.0008 \pm{0.0015}  $  &                  &              \\
9183 -- 9283 & $ 0.0000 \pm{0.0015}  $  &                  &              \\
9283 -- 9383 & $ 0.0019 \pm{0.0022}  $  &                  &              \\
9383 -- 9483 & $-0.0011 \pm{0.0018}  $  &                  &              \\
9483 -- 9578 & $ 0.0035 \pm{0.0021}  $  &                  &              \\
\hline                                                  
\end{tabular}                                        
\end{table*}

\section{Conclusions}
\label{sec:conc}

We present an optical transmission spectrum of the highly irradiated exoplanet WASP-103b using Gemini/GMOS covering the wavelength range between 550 and 960~nm. 
WASP-103b is one of the closest-orbiting hot Jupiters known, with an equilibrium temperature of 2500~K and an orbital separation of less than 1.2 times the Roche limit.
From a combined analysis of observations from three individual transits, we find enhanced absorption at the 589~nm Na and, at lesser significance, at 
the 766~nm K feature. This indicates that the transmission spectrum of WASP-103b is not dominated by strong aerosol absorption, a finding that is in line
with previously-published trends of aerosol appearance and planetary properties \citep{Heng16,Stevenson16}. Based on the altitudes probed by our data, 
we constrain the pressure at the top of any potential clouds in the atmosphere of WASP-103 to be at least 0.01~bar.
At low significance, the Na feature as well as both components of the K feature, show slightly enhanced absorptions compared to 
predictions from a solar-composition planetary atmosphere model. 

Our observations do not confirm a previously-inferred strong trend for increasing absorption at small wavelengths \citep{Southworth15, Southworth16}. 
We have studied the impact of contamination from a blended star on the transmission spectrum, but fail to produce a trend large enough to explain previous observations even 
when assuming errors on the stellar parameters of both stars at the 3-$\sigma$ level.

\begin{acknowledgements}
We thank an anonymous referee for insightful comments that improved the quality of this work. We acknowledge Laetitia Delrez and Catherine Huitson 
for valuable discussions on the treatment of contamination and the wavelength stability of GMOS.
IJ acknowledgements support from the Austrian Research Promotion Agency (FFG) under grant number P847963. Based on observations obtained at the Gemini Observatory, 
which is operated by the Association of Universities for Research in Astronomy, Inc., under a cooperative agreement with the NSF on behalf of the Gemini partnership: 
the National Science Foundation (United States), the National Research Council (Canada), CONICYT (Chile), Ministerio de Ciencia, Tecnolog\'{i}a e Innovaci\'{o}n 
Productiva (Argentina), and Minist\'{e}rio da Ci\^{e}ncia, Tecnologia e Inova\c{c}\~{a}o (Brazil). 

\end{acknowledgements}

\bibliographystyle{aa}
\bibliography{../bbl}

\end{document}